\documentclass[conference,10pt]{IEEEtran}
\usepackage[T1]{fontenc}
\usepackage{amsmath}
\usepackage{subfigure}
\usepackage{graphicx}
\usepackage{textcomp}
\usepackage{floatrow}
\usepackage{algorithm,algpseudocode}
\usepackage{algcompatible}
\usepackage{comment}
\usepackage[left=0.64in,right=0.64in,top=0.7in,bottom=1in]{geometry}
\usepackage{multirow}

\algnewcommand\algorithmicreturn{\textbf{return}}
\algnewcommand\RETURN{\algorithmicreturn}
\usepackage{comment}
\usepackage{amsthm}
\newtheorem{theorem}{Theorem}[]
\newtheorem{definition}{Definition}[]
\begin{document}

\title{ ACE: an Accurate and Cost-Effective Measurement System in SDN }

        
        \author{\IEEEauthorblockN{Fangye Tang, Meysam Shojaee, and Israat Haque}
\IEEEauthorblockA{ Computer Science, Dalhousie University, Halifax, Canada\\
Email: {fangy.tang,meysam.shojaee,israat@dal.ca}}}
        


\maketitle

\begin{abstract}

Packet-level traffic measurement is essential in applications like QoS, traffic engineering, or anomaly detection. Software-Defined Networking (SDN) enables efficient and dynamic network configuration that we can deploy for fine-grained network state measurement. As the state changes dynamically, the sampling frequency must cope with that for accurate measurement. At the same time, we must consider the measurement cost, e.g., nodes' resource utilization. Existing works fall short in offering an optimal model to balance both the measurement accuracy and cost. We fill that gap by proposing  \textit{ACE}, an accurate and cost-effective measurement system that builds on a multi-objective optimization problem. As the optimization problem is NP-hard, we develop a heuristic. We solve the model using CPLEX; then implement a prototype of ACE in Mininet over real-world network topologies. The results confirm that ACE outperforms its counterparts in balancing both accuracy and cost. 


\end{abstract}

\IEEEpeerreviewmaketitle

\section{Introduction}\label{introduction}

Software-Defined Networking (SDN) \cite{ethane, sdn} enables hardware abstraction and network programmability by separating network control logic (control plane) from underlying network elements (data plane) and offers better network configuration and management. The control plane usually deploys the \textit{de facto} OpenFlow protocol \cite{openflow_1.3} to contact the data plane switches. Network administrators can easily configure these by running a predefined script in the controller to apply the configuration to each switch automatically. One of the main advantages of SDN is its flow-based packet forwarding, where each flow has associated counters (e.g., flow duration). Due to the network programmability, the measurement is more convenient in SDN than legacy networks as the administrators can dynamically collect the statistics of the counters from the configured data plane switches on demand. 

Network measurement is essential in applications like QoS provisioning, billing, traffic engineering, anomaly detection, etc. We can deploy two types of measurement in Software-defined Networking (SDN); namely, \textit{active} and \textit{passive} \cite{review}. The active one injects a probe packet into the network, whereas the passive one directly reads the counters based on a sampling rate. Also, there can be a hybrid scheme combining both the active and passive measurements. It can either send probe packets with a fixed polling frequency or read the counters from switches with a fixed sampling rate. For instance, in the case of the measurement tool sFlow\cite{sflow}, a monitoring agent is installed at each switch to send the measured traffic to the monitoring controller. Thus, the collector can poll statistics at a regular interval, or an agent can push these after observing a configured number of packets or flows. 


The polling or sampling frequency, however, has an impact on the measurement accuracy. For active measurement, the higher the polling frequency, the better the accuracy at the price of measurement cost. On the other hand, the higher sampling rate in the passive measurement consumes more resources of the monitored switches. Thus, it is essential to choose an optimal polling or sampling rate. However, existing works mostly rely on a fixed rate and impact both the measurement accuracy and network health. There are a few exceptions where the authors define an on-demand polling/sampling rate. For instance, Payless \cite{payless} proposes a Markov process-based algorithm that uses two consecutive sampling rates to predict the subsequent rate. Sampling-On-Demand (SOD) \cite{sampling_on_demand} designs a model to decide which switches to be sampled at what rate to maximize the measurement accuracy. None of them considers the measurement cost in their design. For instance, the sampling rate at the chosen switches should not overwhelm their CPU resources and degrade performance. Also, switches need to perform their standard task like packet forward or filter; thus, the measurement cost while sampling traffic must be minimized. Therefore, we need a measurement approach that can dynamically assign sampling rates at switches to meet applications' demand on the measurement accuracy while minimizing the cost.  

\textbf{Contribution: }In this paper, we propose an accurate and cost-effective measurement system in SDN called \textit{ACE}. We consider two types of measurements: offline and online, where the latter accommodates new flows on the fly, which is not the case in the offline one. We formulate a multi-objective optimization problem that optimizes both the measurement accuracy and cost. The model selects the sampling switches for the current flows and identifies the corresponding sampling rates while considering the switches' capacity. We prove that the model is NP-hard and develop a greedy heuristic to solve it in a reasonable time. The heuristic allocates flows to the switches to maximize the accuracy and minimize the cost.  

We solve the proposed models in the CPLEX optimizer and provide the model sensitivity analysis. Then, we implement a prototype of ACE using the Ryu SDN controller and a set of Open vSwitches. The switches form two real WAN topologies: USNET and Darkstrand in the Mininet simulator, where Ryu communicates with the switches using OpenFlow protocol. We compare the performance of ACE with two state-of-the-art measurement schemes Payless and SOD. The results reveal that ACE has slightly higher accuracy and saves 50\% and 45\% cost compared to SOD and Payless, respectively.  

The remaining paper structure includes: Section \ref{related} presents related measurement schemes in SDN. We present our models, proof, and the corresponding heuristic in Section \ref{design}. The next section presents the system design, evaluation setup, and discussion on results. We conclude the paper in Section \ref{conclusion}.

\section{Related Work} \label{related}

In this section, we discuss related measurement schemes in SDN. 

\textbf{Active and passive measurements.} In the category of active monitoring, SDN-Mon \cite{sdn-mon} uses a bloom filter to collect flows to be monitored. OpenNetMon (ONM) \cite{opennetmon} polls each flow's destination switch to reduce the resource consumption and improve the throughput, whereas RFlow \cite{rflow} balances between the measurement accuracy and overhead by deploying a set of distributed agents. In the passive monitoring schemes, OpenTM \cite{opentm} is the first traffic matrix estimation scheme in SDN, which periodically samples counters from switches. FlowSense \cite{flowsense} uses OpenFlow \textit{Packet\_in} and \textit{Flow\_Removed} messages to compute flow-lifetime and other statistics. Payless \cite{payless} is a similar OpenFlow-based measurement scheme. The above schemes either need to access every flow or switch, which does not scale, or depend on OpenFlow traffic that can impact the measurement accuracy. 

\textbf{Hybrid measurement.} A set of hybrid schemes are proposed to get the benefit out of both the active and passive measurements. sFlow \cite{sflow} is one such widely used scheme available in most of the switches. sFlow-controller actively polls a set of switches with a fixed polling frequency, and sFlow agents passively sample switches' counters with a fixed sample rate. FlowCover \cite{flowcover} also measures a chosen set of switches, and ReMon \cite{remon} extends it for better measurement cost without degrading the accuracy. CeMon \cite{cemon} is a multi-controller variant of FlowCover, i.e., the measurement task is distributed among a set of controllers. Partial Flow Statistics Collection (PFSC) \cite{pfsc} collects flow statistics from a subset of switches such that the flow recall ratio on every switch reaches a predefined value while minimizing the number of queried switches. Lonely Flow First (LFF) \cite{lff} also monitors a subset of switches. 

\textbf{Determining polling and sampling frequency.} One of the important measurement aspects is to use optimal polling frequency or sampling rate while using an appropriate measurement scheme. Along with this aspect, sFlow uses a static frequency that can impact the accuracy and overhead. Payless \cite{payless} defines a Markov-process-based dynamic polling frequency model that adjusts the frequency based on the OpenFlow statistics. Recently, Oh \textit{et al.} propose a port utilization-based polling frequency determination that can be thought of as a complementary solution to the flow-based ones. OpenSample \cite{opensample} also defines dynamic frequency proportional to the traffic volume. Volley \cite{volley} dynamically adjusts polling frequency based on how likely a state violation can be occurred due to a DDoS attack. Sampling-On-Demand (SOD) \cite{sampling_on_demand} is the state-of-the-art sampling scheme that installs a management module at each switch that the controller configures to indicate which flows to be sampled at which switches and what rates. However, SOD does not consider the measurement cost during such configuration. A similar learning-based sampling frequency prediction is proposed in \cite{netsoft20}, which may suffer from measurement granularity because of the dependency on OpenFlow messages.  We consider both the cost and accuracy while designing the proposed measurement scheme while not confined to OpenFlow messages.  

\section{Optimization Problem}\label{design}

In this section, we present our optimization model for choosing the right sampling rates across switches. The goal of the model is to optimize both the measurement accuracy and cost. Our model has two versions: offline and online, which are presented in Section \ref{sec:offline} and Section \ref{sec:online}, respectively.

\subsection{Offline}\label{sec:offline}
In the offline version, we assume all the flows have already entered the network. Let $S$ be the set of monitoring switches. For each switch $s \in S$, $c_s$ is the sampling capacity of $s$ in packets per second (pps). Also, let $F$ be the set of flows be monitored. For each flow $f \in F$, $r_f$ is the recommended sampling rate. We also assume that there is a computation cost of $C$ for monitoring each flow at a switch, which is proportional to its sampled traffic. Furthermore, $P_f$ is the path of flow $f \in F$. Table \ref{tab1} summarizes our notation.

\newcommand{\minitab}[2][l]{\begin{tabular}{#1}#2\end{tabular}}
\begin{table}[tbp]
\caption{Key notations used in the model.}
\begin{center}
\resizebox{\columnwidth}{!}{
\begin{tabular}{ c c l}
 \hline
& Notation & Description\\
 \hline 
\multirow{4}{*}{Input} & $S$ & set of monitoring switches;\\
& $c_s$ & sampling capacity of switch $s$ in packets per second (pps);\\
& $F$ & set of flows to be monitored;\\
& $r_f$ & recommended sampling rate for flow $f$\\
\hline
\multirow{2}{*}{\minitab[c]{Auxiliary \\ variables}} &  $P_f$ & path of flow $f$; \\
 & $t_{fs}$ & 1 if $s$ in $P_f$, 0 otherwise; \\
 \hline
\multirow{2}{*}{Output} & \( x_{fp} \) & Allocation of flow $f$ on switch $s$;\\
& \( y_{f} \) & sampling rate of flow $f$;\\ 
\hline
\end{tabular}}
\label{tab1}
\end{center}  
\end{table}

\subsubsection{Variables and parameters}
We need to allocate each flow to a switch at which it should be sampled and indicate its sampling rate. For that, we define two sets of decision variables: $x_{fs}$, which is a binary decision variable denoting the switch used for sampling flow $f$, such that
\begin{equation} \label{eq1}
 x_{fs} = \left\{
  \begin{array}{@{}ll@{}}
    1 & \text{if flow $f$ is sampled at switch $s$} \\
    0 & \text{otherwise.} \\
  \end{array}\right. \\
\end{equation}
and continuous decision variable $y_{f}>0$ to  denote  the  sampling  rate  of flow $f$ assigned to switch $s$. We also define binary parameter \( t_{fs} \), to identify the switches belonging to route $P_f$:\\

\begin{equation} \label{eq2}
 t_{fs} = \left\{
  \begin{array}{@{}ll@{}}
    1 & \text{if $s \in P_f$,} \\
    0 & \text{otherwise.} \\
  \end{array}\right. \\
\end{equation}

\subsubsection{Constraints}
In the following, we present the model constraints. 

\textbf{Switch sampling capacity.} capacity constraints (\ref{capacity}) ensure that the total sampling rate required from a switch does not exceed its sampling capacity. The total sampling rate is calculated by summing up the sampling rates of flows allocated to switch $s$.

\begin{equation}
\sum_{f=1}^{|F|} x_{fs} y_{f} \leq c_s, \ \ \text{for each }s \in S \label{capacity}\\    
\end{equation}

\textbf{Flow satisfaction.} constraints (\ref{satisfaction}) ensure each flow is sampled at least at one switch. Note that flow $f$ should traverse a switch $s$ to be sampled at that switch.
\begin{equation}
\sum_{s=1}^{|S|} x_{fs} \geq 1, \ \ \text{for each }f \in F  \label{satisfaction}
\end{equation}


\textbf{Sampling rate.} constraints (\ref{lBound}) ensure sampling rate of a flow is at least the recommended sampling rate. Constraints (\ref{uniqueRate}) also ensure each switch can sample a flow only if the flow traverses that switch.

\begin{equation}
y_f \geq r_f, \ \ \text{for each }f \in F  \label{lBound}
\end{equation}

\begin{equation}
x_{fs}\leq t_{fs}, \ \ \text{for each }f \in F, s\in S   \label{uniqueRate}    
\end{equation}


\subsubsection{Objective function}
The objective function (\ref{objFunc}) consists of two terms. The first term optimizes the accuracy, while the second one reflects the overall resource consumption for the sampling, i.e. the cost. The function also has two parameters, $a$ and $b$, to favor a specific objective depending on operators need (e.g., to favor more accurate sampling due to application needs or lower resource consumption for a better network health).

\begin{equation}
\max \sum_{s=1}^{|S|}\sum_{f=1}^{|F|}  a x_{fs}y_{f} - b Cx_{fs} \label{objFunc}
\end{equation}

\begin{definition} { MC-GAP is an extension of the well-known Knapsack problem called Generalized Assignment Problem (GAP). The input of MC-GAP is a set $B$ of knapsacks and $I$ of items. Each knapsack has a capacity, and each item has a demand and utility. The objective of MC-GAP is to find a feasible assignment of all items to knapsacks within their capacity to maximize the utility. MC-GAP is proved as an NP-hard problem \cite{mc_gap}.}
\end{definition}


\begin{theorem} \label{proof}
The offline sampling model is an NP-hard problem.  
\end{theorem}

\begin{proof}
We can prove the NP-hardness by showing that the proposed offline sampling model is an MC-GAP problem. We can represent each switch as an MC-GAP knapsack whose size is equal to the sampling capacity. Each flow can be represented by an MC-GAP item and each sampling rate as an MC-GAP configuration. The accuracy and cost of sampling a flow at a switch using a sampling rate can be represented by the value in the MC-GAP utility of assigning an item to a knapsack with a configuration. Thus, the offline sampling model is an NP-hard problem.  
\end{proof}

Since the offline model is NP-hard, we propose a greedy heuristic to solve it efficiently based on the heuristic in \cite{mc_gap}. The algorithm is shown in Algorithm \ref{greedy}.

\begin{algorithm}[h]
    \caption{ACE Proposed HeuriStic (APS)}
    \label{greedy}
    \begin{algorithmic}[1]
        \raggedright
        \REQUIRE~~\\
            Monitoring Switches: $S$\\
            Flows: $F$\\
        \STATE $C \gets []$
        \WHILE {$C \neq F$}
            \STATE Find a switch $s \in S$ such that $\sum_{f=1}^{|F|}  a x_{fs}y_{f} - b Cx_{fs}$ is maximum
            \STATE Record $r$ as sampling rate for switch $s$
            \STATE $C \gets C \cup f$
        \ENDWHILE
    \end{algorithmic}
\end{algorithm}

 We assign a score $\sum_{f=1}^{|F|}  a x_{fs}y_{f} - b Cx_{fs}$ to each switch, where the first and second parts are the total accuracy and resource consumption, respectively. Also, we can balance between the accuracy and cost by tuning weights $a$ and $b$, respectively, where $a+b=1$. For instance, their equal value of 0.5 indicates equal importance. However, network operators or designers can easily focus on one feature by increasing its corresponding weight. We greedily choose a switch with a maximum score and record the flows that are sampled in this switch. The process continues until all flows are covered. The main loop iteration in the algorithm takes $\mathcal{O}(n)$, where $n=|F|$. The switch $s$ with the maximum score can be found in $\mathcal{O}(\log m)$ time by a priority queue, where $m=|S|$. Thus, the time complexity of Algorithm \ref{greedy} is $\mathcal{O}(n\log m)$.
 
 \begin{figure}[h]
\centering
\includegraphics[width=0.50\linewidth]{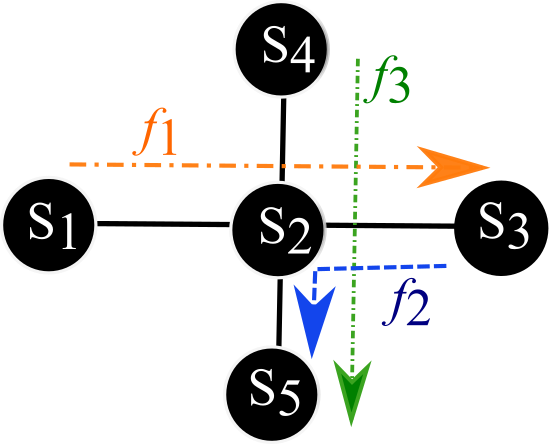}
\caption{Flows $f_1$ and $f_2$ will be sampled at switch $S_2$ while $f_3$ will be sampled at switch $S_5$. Thus, we can balance between the measurement accuracy and cost. }
\label{fig:greedyOperation}
\end{figure} 
 
 We illustrate the operation of Algorithm \ref{greedy} in Fig.~\ref{fig:greedyOperation}. There are three flows \{$f_{1},f_{2}, f_3$\}. Flow $f_{1}$ travels from switch $S_1$ to $S_3$, $f_{2}$ from switch $S_3$ to $S_5$, and $f_{3}$ from switch $S_4$ to $S_5$. We assume that all switches are capable of monitoring flows. There are three switches $S_1$, $S_2$, and $S_3$ at which flow $f_1$ can be sampled. Algorithm \ref{greedy} assigns both flows $f_1$ and $f_2$ to switch $S_2$ while allocates $f_3$ to $S_5$. Note that sampling all three flows at $S_2$ can consume more of its resources despite that it can sample all three flows. Instead, the algorithm distributes the measurement cost among $S_2$ and $S_5$ while keeps the accuracy intact.   

\subsection{Online}\label{sec:online}
In the online version, flows are added one at a time into the network, which can be a new flow or an existing one suffering from a failure. Thus, the controller needs to reconfigure the sampling locations and rates to accommodate the new one. However, without recomputing the entire configuration, we can add a constraint that the sampling rate of an existing flow will only change if the new one is sampled at its associated switch. Let $f'$ be the newly added flow and  $x_{f^{'}s}$ and $y_{f^{'}}$ be the corresponding decision variables. Also, let $F'=F \cup{f'}$ be the set of flows that need to be sampled, where existing flows in $F$ are sampled at the same switches according to the problem in (\ref{capacity})-(\ref{objFunc}). Then, our online model is as follows: 


\begin{align}
& \ \ \ \ \ \ \max \ \ \sum_{s=1}^{|S|} a x_{f^{'}s}y_{f^{'}} - b Cx_{f^{'}s} \label{8} \\
&\text{subject to:} \ \ \sum_{f=1}^{|F'|} x_{fs} y_{f} \leq c_s, \ \ \text{for each }s \in S \label{9}\\
& \hspace{18mm} \sum_{s=1}^{|S|} x_{f's} \geq 1 \ \ \label{10}\\
& \hspace{18mm} y_f' \geq r_f'  \label{11}\\
& \hspace{18mm} x_{f's}\leq t_{f's}, \ \ \text{for each } s\in S  \label{12}
\end{align}

Similar to the offline model, the objective function (\ref{8}) includes both accuracy and resource consumption. Constraints (\ref{9}) ensure that the total sampling rate required from a  switch (for both existing and new flows) does not exceed its sampling capacity. Constraint (\ref{10}) ensures that new flow $f'$ is sampled at least at one switch. Constraint (\ref{11}) ensures the sampling rate of flow $f'$ is at least equal to its recommended sampling rate.  Constraint (\ref{12}) ensures each switch can sample flow $f'$ only if $f'$ traverses that switch.

The online version can also be used when there are link failures. In that case, we can recover from the failures using a solution like ReMon \cite{remon}. After the recovery, we can accommodate those affected flows on the fly using this online scheme. In particular, we can accommodate affected flows in existing switches with the adjusted rate or incorporate new switches. The online formulation is also NP-hard and can be proved following the same proof in \ref{proof}. In the heuristic, we choose switches that can monitor the new flows with the maximum score. The time complexity of the online version is $\mathcal{O}(\log m)$ because we can find switches with the maximum score in a constant time for the constant number of flows. 

\section{ACE Design and Evaluation}\label{discussion}

In this section, we first outline the architecture of ACE. Then, we present the evaluation setup and discussion on model evaluation and simulation results.

\subsection{ACE architecture} 
\begin{figure}[!t]
    \centerline{\includegraphics[width=0.95\textwidth]{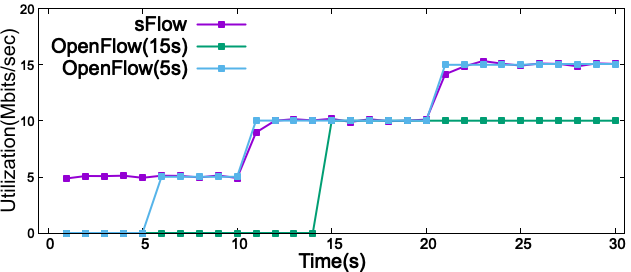}}
    \caption{The measurement granularity of OpenFlow and sFlow. }
    \label{sflow_openflow}
\end{figure}

The proposed greedy heuristic of ACE is deployed as an controller application at the Ryu. It manages a set of programmable switches (OVS) from the data plane. We deploy hybrid measurement scheme sFlow \cite{sflow} that consists of a sFlow-collector (resides at the control plane) and a set of sFlow-agents to offer both the active and passive measurements. The collector gathers the measurement, where sFlow agents installed at the switches samples counters with a rate configured by the ACE heuristic, APS. We consider sFlow instead of OpenFlow statistics as it depends on \textit{flow\_removed} messages, which impacts the accuracy as we observe in Figure~\ref{sflow_openflow}. In this evaluation, we use iperf to generate traffic for five source-destination pairs, each with an initial rate of one Mbps and double it at every ten seconds. Thus, the total network utilization is five Mbps initially. We also set the flow entry's hard timeout to five and fifteen seconds.

\subsection{Evaluation setup} 
\begin{figure}[h]
    \centering
    \subfigure[USNET]{
        \label{usnet}
        \includegraphics[width=0.8\textwidth]{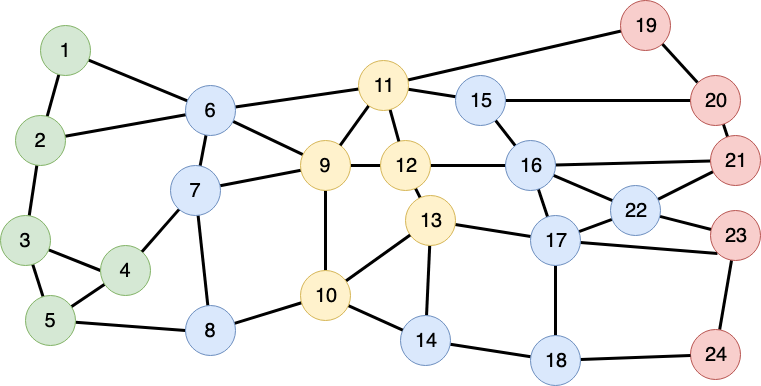}
    }
    \subfigure[Darkstrand]{
        \label{darkstrand}
        \includegraphics[width=0.8\textwidth]{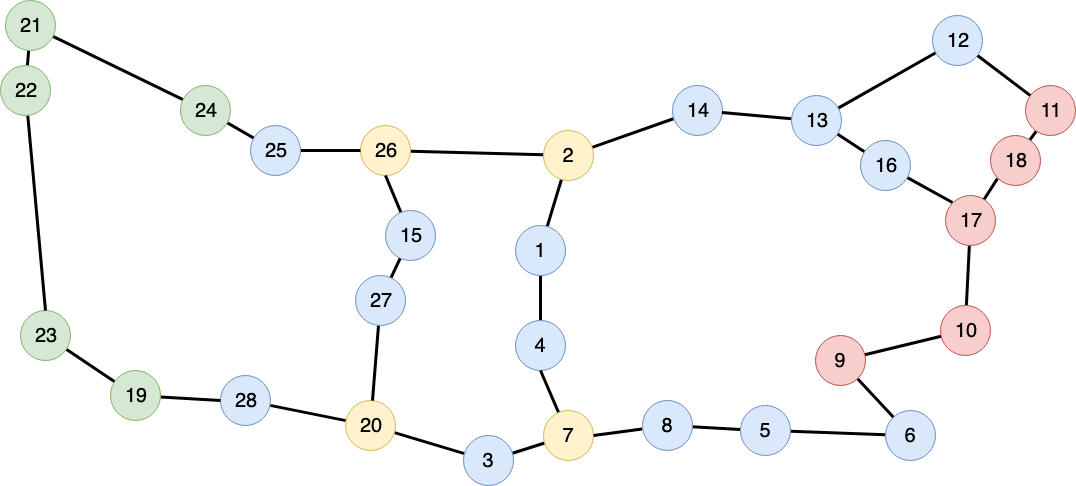}
    }
    \caption{The USNET and Darkstrand topologies.}
    \label{topology}
\end{figure}

We first solve the proposed model using CPLEX. Then, we implement a prototype of ACE in Mininet 2.2.2 emulator resides on a server with an Intel Core i5 2.90 GHz (4 cores) CPU processor and 4GB RAM. The server hosts Ubuntu (64-bit), where we deploy the Ryu \cite{ryu} 4.30 controller to control Open vSwitch \cite{ovs} 2.11.90 virtual switches. Two planes communicate over OpenFlow 1.3 protocol. We consider two real topologies: USNET\cite{usnet} and Darkstrand\cite{darkstrand} in the evaluation. The USNET topology consists of 24 switches and 42 links (Fig.~\ref{usnet}\cite{usnet}) and Darkstrand consists of 28 switches and 31 links (Fig.~\ref{darkstrand}\cite{darkstrand}). We assume each switch is connected with a host, and each host can be a source or destination.

We randomly choose a set of source-destination pairs, where the sources are from green nodes and destinations are from red ones, to force packets to follow the longest routes. Traffic is generated at sources at a rate of 1 Mbps using iperf. We then measure the link utilization using sFlow, where the sampling rate is determined based on the proposed heuristic. The measurement cost is defined as the number of monitoring switches multiplied by their sampling cost. The cost can be defined as CPU resource consumption for the sampling traffic, i.e., the sampling rate; the higher the rate, the larger the cost. 


\subsection{Discussion on Results}

In this section, we first present the model evaluation results from CPLEX. We consider both the USNET and Darstrand topologies to test sensitivity analysis, i.e., the impact of cost vs. accuracy and how to balance that as per the application demand. Each link sets to have a 1\% chance of packet loss. We randomly select 50 source-destination pairs and consider a flow with the rate of 1 Mbps between each pair. We evaluate the effect of variations on the ratio between $a$ and $b$ on accuracy -- the link utilization accuracy computed according to the flow sampling and measurement cost -- the resource consumption at the switches. 

\begin{figure}[h]
\centering
\includegraphics[width=0.9\linewidth]{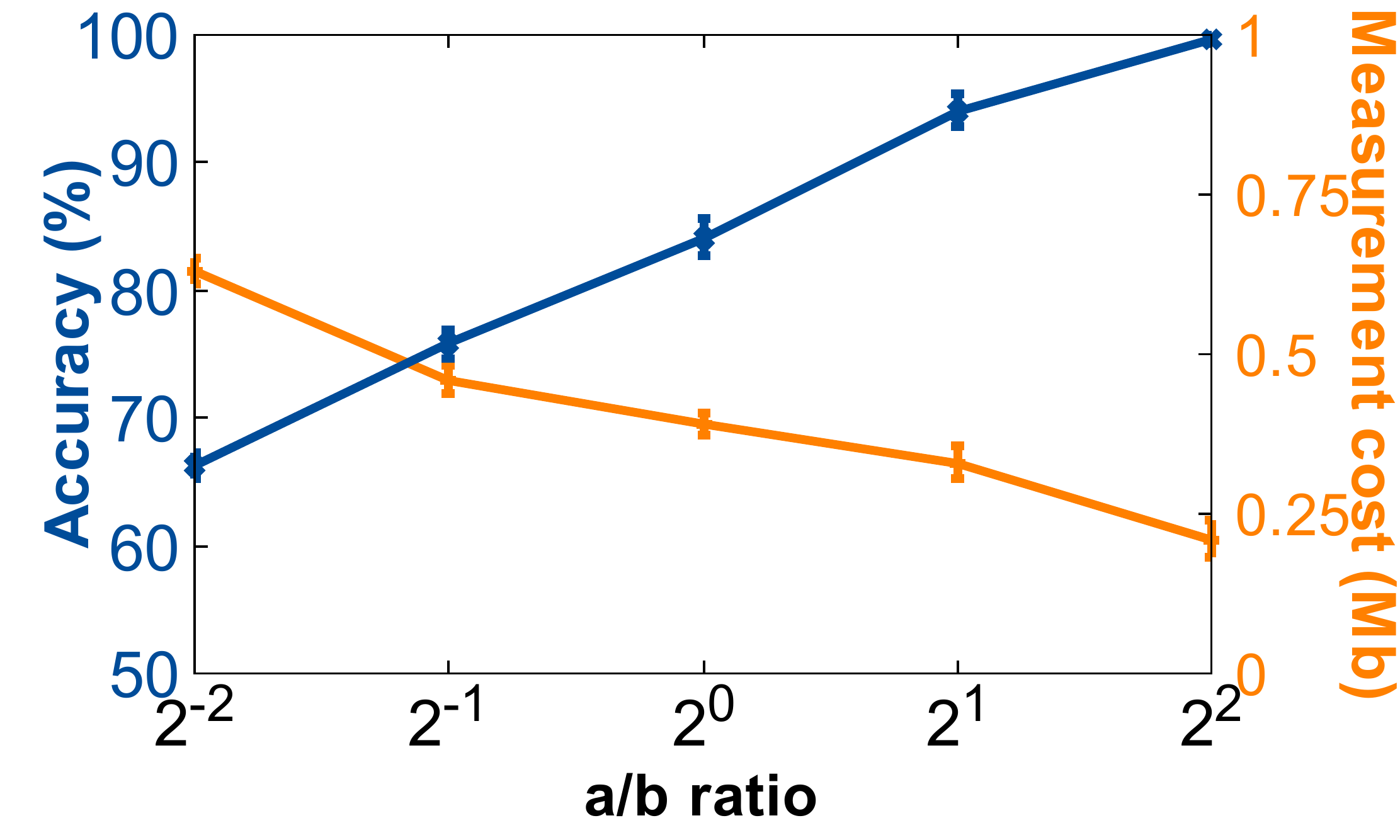}
\caption{Parametric analysis of the model for USNet topology. }
\label{param-sensitivity1}
\end{figure}

\begin{figure}[h]
\centering
\includegraphics[width=0.9\linewidth]{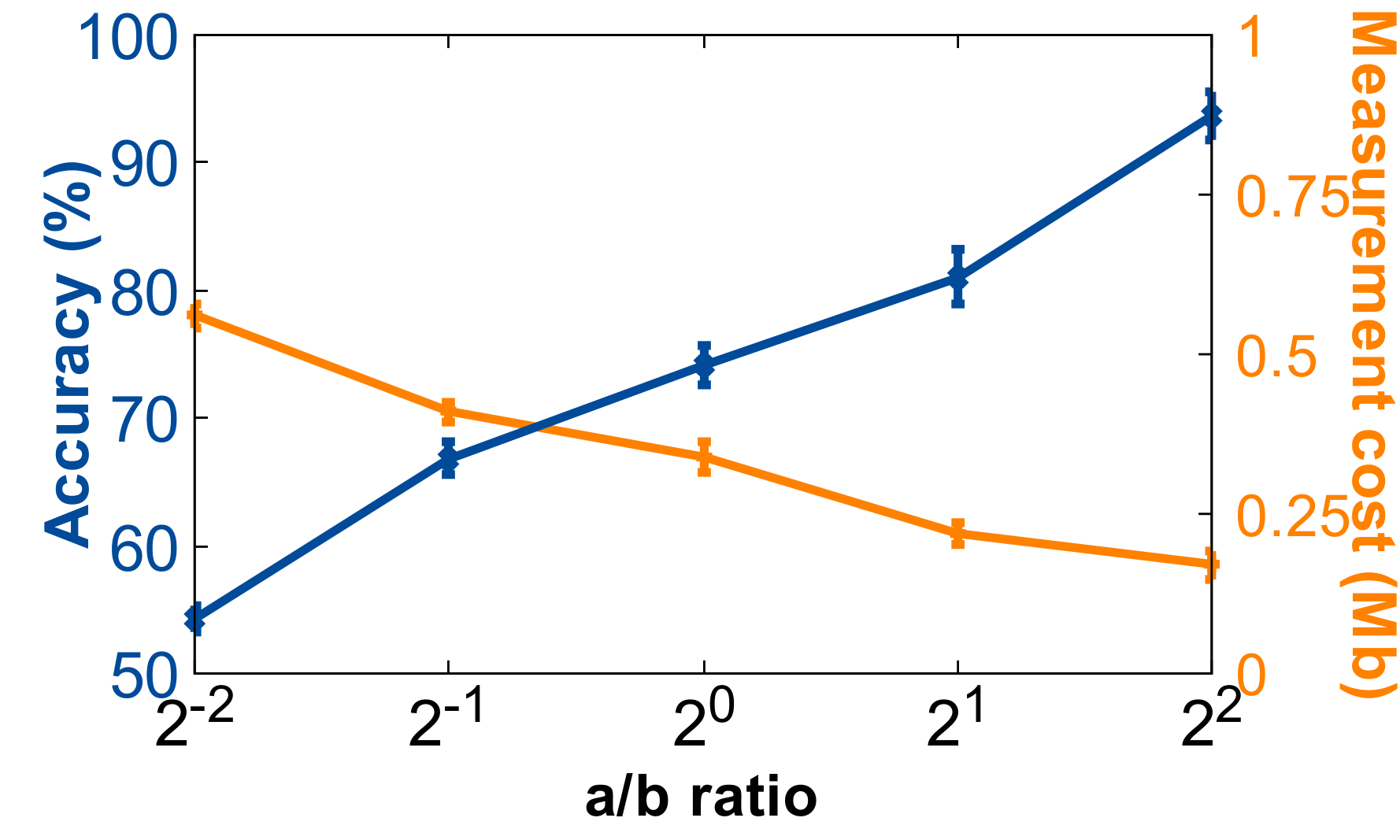}
\caption{Parametric analysis of the model for Darkstrand topology.}
\label{param-sensitivity2}
\end{figure}

Figure \ref{param-sensitivity1} and \ref{param-sensitivity2} show the model evaluation results over USNet and Darkstrand typologies. As expected, in both typologies, the average accuracy increases as we increase the $a/b$ ratio (i.e.,  greater values of $a$ give more importance to the first term of the objective function).  Conversely,  the measurement cost decreases with bigger values of $b$ (i.e., smaller $a/b$ ratios). Also, we see a higher average accuracy and a lower measurement cost in USNet than in Darkstrand topology. The reason is that USNet is denser than Darkstrand, enabling the model to better decide the sampling locations and ratios to achieve a better performance in terms of accuracy and measurement cost.

\begin{figure}[h]
\centering
\includegraphics[width=0.8\linewidth]{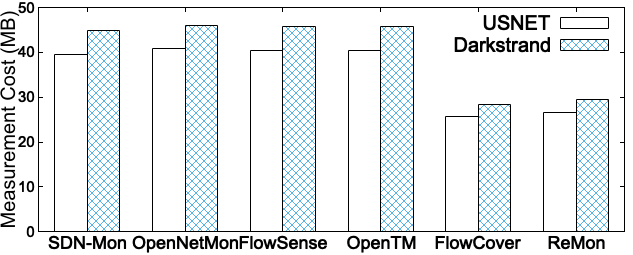}
    \caption{The measurement cost in USNET and Darkstrand. }
    \label{optimization_cost}
\end{figure}

Next, we evaluate the performance of the ACE prototype in the Mininet simulator over the same two topologies. We consider three different a/b ratios: 0.5/0.5 (considering accuracy and cost at the same level), 0.8/0.2 (more weight to the accuracy), and 0.2/0.8(more weight to the cost) to compare the accuracy with Sampling-On-Demand (SOD) (only accuracy). The link utilization measurement accuracy is good and similar in all four schemes, which we do not present due to the space limitation. However, the cost of SOD is high, which is shown in Figure~\ref{optimization_cost}. APS with the ratio of 0.2/0.8 has the lowest cost and obtain a similar accuracy score compared to SOD. 


In the next evaluation, we compare APS with the baseline, payless, and SOD in the Mininet. The baseline is the actual traffic that we generate. The network topology is the USNET, where the chosen sources use iperf to generate traffic for the intended destinations. The sFlow collector gathers the flow statistics from the configured sFlow agents. We dynamically add/remove flows with an initial rate of 1 Mbps and observe the performance of different measurement schemes.

\begin{figure} [h]
    \centerline{\includegraphics[width=0.9\textwidth]{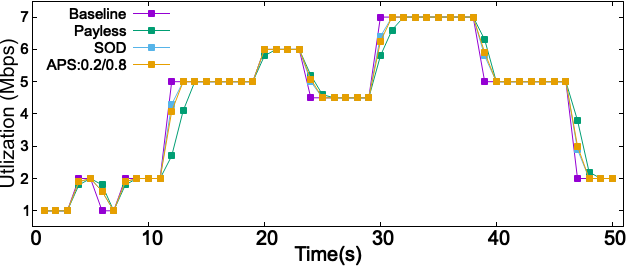}}
    \caption{The measurement accuracy in the USNET topology. }
    \label{optimization_real_accuracy}
\end{figure}

We present the accuracy measures in Fig.~\ref{optimization_real_accuracy}, where in general, all schemes seem to perform similarly; however, we can take a deeper look for some insights. We observe a slight discrepancy in the measurements compared to the baseline when utilization changes, which is the worst for payless and SOD. Another observation is that sFlow also collects some additional control packets like ARP and LLDP packets; thus, the measured traffic can be slightly higher than the actual one. In the controller, we subtract that extra control traffic from the intended measured ones. Note that this computation at the controller is not significant to impact the measurement performance. 


\begin{figure} [h]
    \centerline{\includegraphics[width=0.9\textwidth]{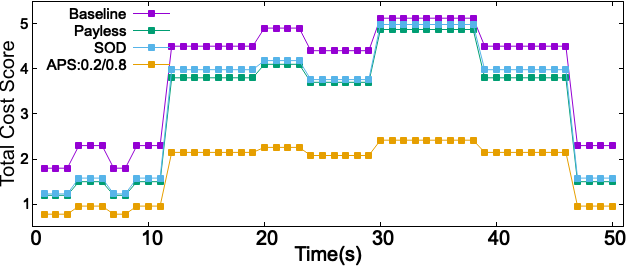}}
    \caption{The measurement cost in the USNET topology. }
    \label{optimization_real_cost}
\end{figure}

On the other hand, Figure~\ref{optimization_real_cost} shows the measurement cost for different schemes, where the baseline approach samples all monitoring switches with a fixed sampling rate. We consider APS with a ratio of 0.2/0.8 as the focus here is to show the measurement cost. APS and the baseline have the best and worst performance, respectively, as expected. Payless and SOD are in between and have almost the same performance, with payless is slightly better. Interestingly, the discrepancy between APS and other schemes is not large at low load, which is significant when network load increases. In particular, we observe around 50\% measurement cost reduction in the case of APS compared to SOD and Payless.

\section{Conclusion}\label{conclusion}
We have proposed an accurate and cost effective measurement system called ACE. In its design, we have first model a multi-objective problem to optimization measurement accuracy and cost. As the problem is NP-hard, we have developed a greedy heuristic to solve the problem at scale. The model is evaluated in CPLEX; furthermore, we have implemented a prototype of ACE in Mininet over two real topologies having varying densities, where the heuristic runs on the Ryu SDN controller. The evaluation results have indicated that ACE can reduce 50\% measurement cost and maintained the same level of accuracy compared to the state-of-the-art design SOD.



\ifCLASSOPTIONcompsoc
  \section*{Acknowledgments}
\else
\fi


\ifCLASSOPTIONcaptionsoff
  \newpage
\fi

\bibliographystyle{IEEEtran}
\bibliography{reference}

\begin{IEEEbiography}{Michael Shell}
Biography text here.
\end{IEEEbiography}

\begin{IEEEbiographynophoto}{John Doe}
Biography text here.
\end{IEEEbiographynophoto}

\begin{IEEEbiographynophoto}{Jane Doe}
Biography text here.
\end{IEEEbiographynophoto}

\end{document}